\documentclass[11pt]{article}
\usepackage{fullpage,epsf,amssymb,amsthm,amsfonts,amsmath,latexsym,array,graphicx,extarrows,mathtools,mathstyle,mathrsfs,slashed,subcaption,amsbsy,csquotes,accents}
\usepackage[normalem]{ulem}
\usepackage{authblk}
\usepackage{cite}
\usepackage{color}    
\usepackage{float} 
\usepackage{hyperref}
\usepackage[export]{adjustbox}
\usepackage{hhline}

\numberwithin{equation}{section}

\newtheorem{theorem}{Conclusion}

\title{\bf{Goldstone bosons across thermal phase transitions}} 

\author{Peter Lowdon}
\author{Owe Philipsen}

\affil{{\scriptsize Institut f\"{u}r Theoretische Physik, Goethe-Universit\"{a}t, Max-von-Laue-Str. 1,  60438 Frankfurt am Main, Germany}}

\date{}
\begin{document}
\maketitle

\begin{abstract}
\noindent
Temperature has a significant effect on the properties of quantum field theories (QFTs) with a spontaneously broken symmetry, in particular on the massless Goldstone bosons that exist in the vacuum state. It has recently been shown using lattice calculations for a $\mathrm{U}(1)$ complex scalar field theory that the Goldstone mode persists even when the symmetry is restored above the critical temperature $T_{c}$, and has the properties of a screened excitation, a so-called thermoparticle. In this work, we continue the investigation of this theory by determining explicitly how the Goldstone mode evolves as the temperature is increased both below and above $T_{c}$. We find that the two phases of the theory are entirely characterised by the thermal dissipative effects experienced by the Goldstone mode, with the broken and symmetry-restored phases associated with weak and strong damping, respectively. These findings are consistent with the non-perturbative constraints imposed by spontaneous symmetry breaking, and provide a new way in which to characterise thermal phase transitions in QFTs. 
\end{abstract}

\newpage       

\section{Introduction}

For many years it has been understood that if the Hamiltonian of a theory is invariant under a continuous global symmetry, and this symmetry is spontaneously broken, which means that the ground state of the system is not symmetric, then there must exist a state whose energy $\omega(\vec{p})$ vanishes in the zero-momentum limit $\vec{p}\rightarrow 0$. This is the main implication of Goldstone's theorem~\cite{Goldstone:1962es,Kastler:1966wdu}. In the case of relativistic systems, this state satisfies $p^{2}=0$ and becomes a stable massless particle, a Goldstone boson. The mechanism by which continuous global symmetries are spontaneously broken lies at the heart of many important physical phenomena, including the existence of collective modes in condensed matter systems, phase transitions, and the interplay of different interactions within the Standard Model of particle physics~\cite{Strocchi:2021abr}. \\ 

\noindent
For non-vanishing temperatures $T>0$ there still remain many open questions regarding Goldstone's theorem and the effects of spontaneous symmetry breaking. In particular, what happens to the Goldstone modes that occur in the vacuum theory, and how do they change if phase transitions exist at high $T$?~\cite{Kirzhnits:1972ut,Dolan:1973qd,Weinberg:1974hy}. Understanding these characteristics is essential for describing the dynamics of systems such as the early universe, and nuclear matter in extreme environments. When $T>0$ the boost invariance of the ground state of a relativistic system is lost, and Goldstone's theorem no longer implies the existence of on-shell massless Goldstone modes. Nevertheless, many important non-perturbative characteristics remain~\cite{Bros:1996kd}. The goal of this work is to carefully outline the non-perturbative aspects of Goldstone's theorem in quantum field theories (QFTs) at both $T=0$ and $T>0$, and use these results to continue the investigation in Ref.~\cite{Lowdon:2025fyb} of the Goldstone mode in the $\mathrm{U}(1)$ complex scalar theory, in particular how it evolves as a function of $T$, and what this means for the phase structure of the theory. The remainder of the paper is organised as follows: in Sec.~\ref{Goldtone_QFT} we review the non-perturbative formulation of Goldstone's theorem and its main implications in vacuum and at finite temperature, in Sec.~\ref{goldstone_u1} we analyse lattice data of the $\mathrm{U}(1)$ complex scalar theory and discuss the physical implications in the context of the conclusions from Sec.~\ref{Goldtone_QFT}, and in Sec.~\ref{concl} we summarise our findings.

\section{Goldstone's theorem for QFTs}
\label{Goldtone_QFT}

Goldstone's theorem establishes the spectral constraints that spontaneous symmetry breaking imposes on quantum systems. In this section we discuss how this theorem is rigorously formulated for QFTs in vacuum, and its generalisation to non-zero temperatures. We review in particular the analysis of Ref.~\cite{Bros:1996kd}, detailing the key derivations and their implications for symmetry restoration at finite temperature.

\subsection{Zero temperature}
\label{GoldstoneT0}

If the dynamics of a QFT are symmetric with respect to a continuous global symmetry which is preserved under spacetime translations, resulting in a local\footnote{A local conserved current satisfies $\partial^{\mu}j_{\mu}=0$ and the condition: $\left[j_{\mu}(x),\phi(y)\right]=0$ for $(x-y)^{2}<0$. The latter property is assumed to hold for all fields in the theory, which guarantees that space-like separated measurements commute with one another and therefore preserve causality~\cite{Streater:1989vi,Haag:1992hx,Bogolyubov:1990kw}.} conserved current $j_{\mu}$, Goldstone's theorem implies that if the ground state of the system is not invariant under the transformations generated by $j_{\mu}$ then this cannot represent a full symmetry of the system~\cite{Goldstone:1962es,Kastler:1966wdu}. In this sense, the symmetry is said to be spontaneously broken. For classical theories, the conservation of the current and absence of boundary contributions is sufficient to guarantee that the charge $Q = \int d^{3}\vec{x} \, j_{0}(t,\vec{x})$ is time independent and generates the corresponding symmetry. However, the direct extension of this definition to quantum systems fails because in quantum theories the current operator $j_{0}(t,\vec{x})$ is a distribution, which in general cannot be integrated\footnote{For a comprehensive discussion on the distributional nature of quantum fields see Refs.~\cite{Streater:1989vi,Haag:1992hx,Bogolyubov:1990kw}.}, and hence the classical definition of $Q$ is ill defined~\cite{Lopuszanski:1991,Strocchi:2021abr}. \\

\noindent
In order to resolve this issue, one must instead define a \textit{localised} charge operator $Q_{\delta R}$ by integrating the current with functions which are non-vanishing over some finite region of space $|\vec{x}| \leq R$, and time $|t|\leq \delta$. In particular, this can be defined as~\cite{Lopuszanski:1991,Strocchi:2021abr}  
\begin{align}
Q_{\delta R} = \int d^{4}x \, \alpha_{\delta}(t)g_{R}(\vec{x})j_{0}(t,\vec{x}), \quad   g_{R}(\vec{x}) = g\left(|\vec{x}|/R\right), \ \ g(x) = \left\{ 
\begin{array}{cc}
      1, & |x|\leq 1 \\
      0, & |x|> 1
    \end{array}   \right. 
\label{Q_def}
\end{align}
where $\alpha_{\delta}(t)$ has compact support and is such that $\alpha_{\delta}(t) \xrightarrow[]{\delta \rightarrow 0} \delta(t)$. This regularised charge approaches the naive classical definition in the $R\rightarrow \infty$ and $\delta \rightarrow 0$ limits, and represents a well-defined QFT operator for any finite value of $R$ and $\delta$. However, since the explicit limits $\lim_{\delta \rightarrow 0}\lim_{R \rightarrow \infty} Q_{\delta R}$ do not converge, it is important to understand precisely how the behaviour of $Q_{\delta R}$ is connected to whether the symmetry is realised or not. It turns out that a necessary and sufficient condition for the symmetry to exist, and be generated by a charge operator $Q$ via $\delta A = i[Q,A]$, is that every local field $A$ must satisfy\cite{Lopuszanski:1991} 
\begin{align}
\lim_{R \rightarrow \infty} \langle  [Q_{\delta R},A ] \rangle_{0} = 0,
\label{SSB_vev}
\end{align}
where $\langle \, \cdot \, \rangle_{0}$ denotes the vacuum expectation value. Taking the converse of this condition, it immediately follows that if \textit{any} operator $A$ exists such that the above limit is non-vanishing, the symmetry must be spontaneously broken. The power of Eq.~\eqref{SSB_vev} is that the limit is guaranteed to be well-defined independently of whether $Q$ exists or not~\cite{Lopuszanski:1991}. This avoids potential contradictory statements about the occurrence of spontaneous symmetry breaking which require $Q$ to act non-trivially on the ground state, even though its existence is not guaranteed. It also makes clear that this phenomenon is connected to the large-distance quantum fluctuations of the ground state. Goldstone's theorem has significant consequences for the spectral properties of QFTs~\cite{Strocchi:2021abr}:
\begin{theorem}
Given an operator $A$ for which $\lim_{R \rightarrow \infty} \langle [Q_{\delta R},A]  \rangle_{0} \neq 0$, the Fourier transform of the correlation function $\langle [j_{0}(x), A(y)]\rangle_{0}$ must contain a $\delta(p^{2})$ singularity.
\end{theorem}
As is well-known, this singularity represents the existence of a stable massless particle state in the spectrum, a Goldstone boson.

\subsection{Finite-temperature generalisation}
\label{TGen}

A non-zero temperature $T=1/\beta>0$ immediately affects the conclusions of Goldstone's theorem, including the existence of a stable massless Goldstone state. Nevertheless, much of the analysis in the zero-temperature case can be generalised, including the condition in Eq.~\eqref{SSB_vev}, except now the expectation value is taken with respect to the thermal ground state $\langle \, \cdot \, \rangle_{\beta}$. For a scalar field theory it follows from the locality and conservation properties of $j_{\mu}$ that 
\begin{align}
\lim_{R\rightarrow \infty} \langle[Q_{\delta R}, \phi(0)]\rangle_{\beta} = q \int_{-\infty}^{\infty} \! dt \, \alpha_{\delta}(t),
\label{SSB_q}
\end{align}
where $q$ is a complex number independent of the specific choice of $\alpha_{\delta}$ and $g$ in the definition of $Q_{\delta R}$. Equation~\eqref{SSB_q} represents the quantum generalisation of the time-independence of the classical charge. Combining Eqs.~\eqref{Q_def} and~\eqref{SSB_q} it follows that the spectral function 
$\rho_{j_{0}\phi}(\omega,\vec{p})$, defined as 
\begin{align}
\rho_{j_{0}\phi}(\omega,\vec{p}) = \int \! d^{4}x \ e^{i p \cdot x} \, \langle[j_{0}(x), \phi(0)]\rangle_{\beta},
\end{align}
satisfies the condition 
\begin{align}
&\lim_{R\rightarrow \infty} \int_{-\infty}^{\infty} \frac{d\omega}{2\pi} \int \!\frac{d^{3}\vec{p}}{(2\pi)^{3}} \, \tilde{\alpha}_{\delta}(\omega) \tilde{g}(\vec{p}) \,\rho_{j_{0}\phi}(\omega,\vec{p}/R) = q \tilde{\alpha}_{\delta}(0),
\label{rho_SSB}
\end{align}
which after taking the $R\rightarrow \infty$ limit, and using the condition $\int \frac{d^{3}\vec{p}}{(2\pi)^{3}}\tilde{g}(\vec{p})=1$, implies 
\begin{align}
\int_{-\infty}^{\infty} \frac{d\omega}{2\pi}  \tilde{\alpha}_{\delta}(\omega)  \,\rho_{j_{0}\phi}(\omega,\vec{p}=0) = q \tilde{\alpha}_{\delta}(0).
\label{rho_SSB2}
\end{align}
In the broken phase $q \neq 0$ one is then led to the conclusion
\begin{align}
\rho_{j_{0}\phi}(\omega,\vec{p}=0) = 2\pi q \, \delta(\omega),
\end{align}    
which agrees with the well-known result in the vacuum theory that the spectrum contains a zero-energy excitation in the $\vec{p}\rightarrow 0$ limit~\cite{Strocchi:2021abr}. \\ 

\noindent
An important question is whether thermal Goldstone modes possess other distinctive characteristics, in particular for non-vanishing momenta. A significant breakthrough in this regard was made in Ref.~\cite{Bros:1996kd}, which used the fundamental constraints imposed by causality, namely that the current-field commutator satisfies: $\left[j_{\mu}(x),\phi(y)\right]=0$, for $(x-y)^{2}<0$. In previous work by the same authors~\cite{Bros:1992ey,Bros:1996mw} it was demonstrated that this condition implies that the Fourier transform of causal commutators must satisfy a non-perturbative spectral representation\footnote{For a recent discussion of this representation see Ref.~\cite{Nair:2025jgl}.}. In the present case this means that $\rho_{j_{0}\phi}(\omega,\vec{p})$ can be written in the general form~\cite{Bros:1996kd}
\begin{align}
\rho_{j_{0}\phi}(\omega,\vec{p}) = \int_{0}^{\infty} \! ds \int \! \frac{d^{3}\vec{u}}{(2\pi)^{2}} \ \epsilon(\omega) \, \delta\!\left(\omega^{2} - (\vec{p}-\vec{u})^{2} - s \right) \left[ -i\omega \widetilde{D}_{\beta}^{(+)}(\vec{u},s) + \widetilde{D}_{\beta}^{(-)}(\vec{u},s) \right].
\label{spec_jphi}
\end{align}  
Equation~\eqref{spec_jphi} corresponds to the $T>0$ generalisation of the well-known K\"{a}ll\'{e}n-Lehmann representation for QFTs at $T=0$~\cite{Kallen:1952zz,Lehmann:1954xi}. A significant implication of Eq.~\eqref{spec_jphi} is that the behaviour of $\rho_{j_{0}\phi}(\omega,\vec{p})$ is entirely fixed by $\widetilde{D}_{\beta}^{(\pm)}(\vec{u},s)$. These thermal spectral densities therefore hold the key to determining the type of excitations that can exist when $T>0$. Since the right-hand-side of Eq.~\eqref{SSB_q} vanishes for odd functions of $t$, one can restrict to the specific case where the temporal function is even, and for simplicity, has unit normalisation. One can also further assume this function to have the spacelike time-averaged form $\alpha_{\delta R}(t)=\alpha\!\left(t/\delta R\right)/\delta R$ with $\alpha(t)$ of compact support, as used for example in non-perturbative analyses of the Higgs mechanism~\cite{Morchio:2006vj}. With this form $\alpha_{\delta R}(t)$ automatically satisfies $\alpha_{\delta R}(t) \xrightarrow[]{\delta \rightarrow 0} \delta(t)$, and the normalisation condition implies $\int_{-\infty}^{\infty} \! dt \, \alpha(t)=1$. 

\subsubsection{Vacuum Goldstone mode} 
\label{vac_goldstone}

Using the spectral representation in Eq.~\eqref{spec_jphi}, together with the definition of the temporal function $\alpha_{\delta R}(t)$, the spontaneous symmetry breaking condition in Eq.~\eqref{rho_SSB} reduces to   
\begin{align}
\lim_{\delta \rightarrow 0}\lim_{R\rightarrow \infty}\int_{0}^{\infty} \!\! ds \int \! \frac{d^{3}\vec{u}}{(2\pi)^{3}} \frac{d^{3}\vec{p}}{(2\pi)^{3}} \ \widetilde{D}_{\beta}^{(+)}(\vec{u},s) \, \tilde{g}(\vec{p}) \,  \tilde{\alpha}\!\left(\delta R\sqrt{\left((\vec{p}/R)-\vec{u}\right)^{2} + s}\right)  = iq \tilde{\alpha}(0),
\label{SSB_damping}
\end{align}
where the dependence on $\widetilde{D}_{\beta}^{(-)}$ drops out due to the evenness of $\alpha_{\delta R}$, and the integrand now involves the Fourier transform $\tilde{g}$, $\tilde{\alpha}$ of the spatial and temporal functions. At zero temperature, a further constraint is introduced by the so-called \textit{spectral condition}, which assumes that all states must have positive energy and $p^{2} \geq 0$~\cite{Streater:1989vi,Haag:1992hx,Bogolyubov:1990kw}. This ultimately implies that $\widetilde{D}_{\beta}^{(+)}(\vec{u},s)$ must be proportional to $\delta^{3}(\vec{u})$, and in particular one can define 
\begin{align}
\widetilde{D}_{\beta}^{(+)}(\vec{u},s)= iq(2\pi)^{3}\delta^{3}(\vec{u})\rho^{(+)}(s),
\end{align}
where $\rho^{(+)}(s)$ is the vacuum spectral density. Equation~\eqref{SSB_damping} then becomes
\begin{align}
\lim_{\delta \rightarrow 0}\lim_{R\rightarrow \infty}\int_{0}^{\infty} \!\! ds  \int \frac{d^{3}\vec{p}}{(2\pi)^{3}} \ \rho^{(+)}(s) \, \tilde{g}(\vec{p}) \,  \tilde{\alpha}\!\left(\delta R\sqrt{(\vec{p}/R)^{2} + s}\right)  = \tilde{\alpha}(0).
\label{SSB_damping_T0}
\end{align}
Any spectral component with $s>0$ must give a vanishing contribution, since one can take the limit $R\rightarrow \infty$ inside the integral\footnote{This follows from the dominated convergence theorem~\cite{Strocchi:2021abr}.}, and use the fact that
\begin{align}
\lim_{R\rightarrow \infty}\tilde{\alpha}\!\left(\delta R\sqrt{(\vec{p}/R)^{2} + s}\right) = \lim_{R\rightarrow \infty}\tilde{\alpha}\!\left(\delta R \sqrt{s}\right)=0,
\end{align}
where the last equality holds because $\tilde{\alpha}(\omega)$ vanishes for $\omega\rightarrow \infty$ on account of $\alpha(t)$ having compact support. From this argument it follows that $\rho^{(+)}(s)$ can only contribute at $s=0$, in particular: $\rho^{(+)}(s) = \delta(s)$, which means that $\rho_{j_{0}\phi}(\omega,\vec{p})$ must contain a $\delta(p^{2})$ component. This is simply Conclusion~1 of Sec.~\ref{GoldstoneT0}. From this analysis one can see that the additional physical constraints imposed at $T=0$ causes the spectral function condition $\rho_{j_{0}\phi}(\omega,\vec{p}=0) = 2\pi q \, \delta(\omega)$ in the broken phase for $T>0$ to be extended beyond $\vec{p}=0$ onto the mass shell $p^{2}=0$, and hence the Goldstone mode in vacuum becomes a stable massless particle state as $T\rightarrow 0$. 

\subsubsection{Thermal Goldstone mode}  
\label{T_goldstone}
 
Since Eq.~\eqref{SSB_damping} reproduces the well-known spectral consequences of Goldstone's theorem at $T=0$, an important question is whether one can use this constraint to infer additional information about the Goldstone mode for $T>0$. In Ref.~\cite{Bros:1996kd} the authors observed that in this case Eq.~\eqref{SSB_damping} can be rewritten as follows 
\begin{align}
q &= -i\lim_{\delta \rightarrow 0}\lim_{R\rightarrow \infty}\int_{0}^{\infty} \!\! ds \int \! \frac{d^{3}\vec{u}}{(2\pi)^{3}} \ \widetilde{D}_{\beta}^{(+)}(\vec{u},s) \,  \tilde{\alpha}\!\left(\delta R\sqrt{|\vec{u}|^{2} + s}\right) \nonumber \\
&= -i\lim_{\delta \rightarrow 0}\lim_{R\rightarrow \infty}\int_{0}^{\infty} \!\! ds \int \! d^{3}\vec{x} \ D_{\beta}^{(+)}\!\left(\delta R \,\vec{x},s \right) \int \! \frac{d^{3}\vec{u}}{(2\pi)^{3}}  \,  e^{-i\vec{u}\cdot \vec{x}} \tilde{\alpha}\!\left(\sqrt{|\vec{u}|^{2} + (\delta R)^{2}s}\right),
\label{SSBq}
\end{align}
which makes use of the fact that the normalisation condition for $\alpha(t)$ and definition of $g_{R}(\vec{x})$ imply $\tilde{\alpha}(0)=1$ and $\int \frac{d^{3}\vec{p}}{(2\pi)^{3}}\tilde{g}(\vec{p})=1$, respectively. Technically, Eq.~\eqref{SSBq} generalises the relation derived in Ref.~\cite{Bros:1996kd}, since there the authors made the specific choice $\tilde{\alpha}(\omega)= e^{-\lambda^{2}\omega^{2}/2}$ for the temporal function, which is consistent with our regularisation when $\lambda =\delta R$. By writing Eq.~\eqref{SSB_damping} in this way, this emphasises a crucial property  
\begin{theorem}
Whether a symmetry is spontaneously broken $(q\neq 0)$ or restored $(q=0)$ at finite temperature depends on the properties of the thermal spectral density $D^{(+)}_{\beta}(\vec{x},s)$.
\end{theorem}
This confirms the physically intuitive picture that the non-perturbative dynamics of the thermal medium are responsible for whether continuous symmetries persist or are screened when $T>0$. \\

\noindent
Under the condition that $D_{\beta}^{(+)}(\vec{x},s)$ decreases monotonically in $|\vec{x}|$, and defines a measure in $s$, as it does at zero temperature\footnote{For $D_{\beta}^{(+)}(\vec{x},s)$ to define a measure this requires that it must be well defined when integrated with any continuous function of $s$. At zero temperature, this follows from the states having a positive Hilbert space norm~\cite{Bogolyubov:1990kw}.}, in Ref.~\cite{Bros:1996kd} the authors were able to further prove that as long as $D_{\beta}^{(+)}(\vec{x},s)$ does not vanish for asymptotically large $|\vec{x}|$, it must contain a discrete contribution of the form $D^{(+)}_{\beta}(\vec{x})\delta(s)$ when $q \neq 0$, and hence has the following decomposition\footnote{The pre-factors of $i$ arise from the definition $\delta A = i[Q,A]$.}
\begin{align}
D_{\beta}^{(+)}(\vec{x},s) = iD^{(+)}_{\beta}(\vec{x})\delta(s) + i D^{(+)}_{\beta,r}(\vec{x},s),
\label{D_decomp}
\end{align}
where $D^{(+)}_{\beta,r}(\vec{x},s)$ contains the remainder of the spectral contributions, which are not concentrated at $s=0$. In the zero-temperature limit $D^{(+)}_{\beta}(\vec{x})\delta(s) \rightarrow q\,\delta(s)$, and hence this particle-like component represents the thermal generalisation of the vacuum Goldstone boson. The function $D^{(+)}_{\beta}(\vec{x})$ has the physical interpretation of a thermal damping factor, since $D^{(+)}_{\beta}(\vec{x})$ having a non-trivial structure implies that the massless Goldstone peak at $p^{2}=0$ becomes broadened, and its overall amplitude is reduced, which captures the dissipative effects that the Goldstone mode experiences as it moves through the thermal medium. The structure described by Eq.~\eqref{D_decomp} is actually the massless realisation of a proposition $D_{\beta}(\vec{x})\delta(s-m^{2})$ first put forward in Ref.~\cite{Bros:1992ey} for how stable particle states of mass $m$ should contribute to the thermal spectral density when $T>0$. These components were later referred to as \textit{thermoparticles}~\cite{Buchholz:1993kp}. \\

\noindent
Based on the analytic properties of the spectral representation in Eq.~\eqref{SSBq}, the authors of Ref.~\cite{Bros:1996kd} pointed out a further remarkable possibility:
\begin{theorem}
If the symmetry is thermally restored, and hence $q=0$, the thermal Goldstone component $D^{(+)}_{\beta}(\vec{x})\delta(s)$ can still give a non-trivial contribution.
\end{theorem}  
This implies that the thermal Goldstone mode need not cease to exist, even in the symmetry-restored high-temperature phase. The observation of a such a thermal Goldstone mode above $T_{c}$ was recently confirmed by lattice studies of the complex scalar $\mathrm{U}(1)$ model in Ref.~\cite{Lowdon:2025fyb}, which strongly suggests that the decomposition in Eq.~\eqref{D_decomp} holds at \textit{any} temperature, independently of whether one is in the symmetry-broken or restored phase. Upon substituting this decomposition into Eq.~\eqref{SSBq}, and using the same argument as in the zero-temperature case in Sec.~\ref{vac_goldstone}, one finds that the component $D^{(+)}_{\beta,r}(\vec{x},s)$ gives no contribution to the integral due to its non-vanishing behaviour when $s>0$. The phase of the theory is therefore determined entirely by the thermal Goldstone component, in particular 
\begin{align}
q = -\lim_{\delta \rightarrow 0}\lim_{R\rightarrow \infty} \int \! \frac{d^{3}\vec{x}}{(2\pi)|\vec{x}|} \ D^{(+)}_{\beta}(\delta R \, \vec{x}) \,  \dot{\alpha}(|\vec{x}|) = \lim_{|\vec{x}|\rightarrow \infty} \! D^{(+)}_{\beta}(\vec{x}),
\label{q_lim_TP}
\end{align}
where the final equality follows from the normalisation condition for $\alpha(t)$. Equation~\eqref{q_lim_TP} immediately implies the following characterisation of spontaneous symmetry breaking at finite temperature:
\begin{theorem}
Spontaneous symmetry breaking and restoration at finite temperature is determined by the asymptotic damping of the Goldstone mode. If the mode experiences: 
\begin{align*}
&\text{(i) Weak dissipation}, \ \lim_{|\vec{x}|\rightarrow \infty} D^{(+)}_{\beta}(\vec{x}) \neq 0, \ \text{the system is in the broken phase}  \\
&\text{(ii) Strong dissipation}, \ \lim_{|\vec{x}|\rightarrow \infty} D^{(+)}_{\beta}(\vec{x}) = 0, \ \text{the system is in the symmetry-restored phase} 
\end{align*}
\end{theorem}  

These conditions are in line with the physical picture that dissipative effects at sufficiently high temperatures can destroy the long-range order of the system, causing the symmetry to be restored~\cite{Bros:1996kd}. \\

\noindent
So far this analysis has been model independent, and $q$ represents the order parameter of the system. In the specific case of a complex scalar theory $q=\langle \phi \rangle_{\beta}$\footnote{In general this quantity is complex, but one can always choose a rescaling of the field operator $\phi(x)$ in which the vacuum expectation value is purely real. Different choices of rescaling lead to physically equivalent, but unitarily inequivalent, representations~\cite{Nakanishi:1990qm}.}. The analysis above makes it clear that the behaviour of the order parameter is controlled by the damping factor of the thermal Goldstone mode, in particular its large-distance behaviour. How this evolves as a function of temperature, and ultimately the nature of any thermal phase transition, is therefore determined by the dissipative effects experienced by this state as it moves through the medium, which is a consequence of the underlying dynamics of the theory. The phase boundary is characterised by a \textit{discontinuity} in these dissipative effects for temperatures below and above $T_{c}$. Another important consequence of Eq.~\eqref{q_lim_TP} is that symmetry restoration only requires $D^{(+)}_{\beta}(\vec{x})$ to vanish in the $|\vec{x}|\rightarrow \infty$ limit, but does not specify the rate at which this might occur. Therefore, in principle if $D^{(+)}_{\beta}(\vec{x}) \sim |\vec{x}|^{-\varepsilon}$ with $\varepsilon>0$ at large distances, this is sufficient to guarantee that the symmetry is restored. This results in an interesting possibility: 
\begin{theorem}
High-temperature symmetry restoration can occur without requiring the correlation functions to have an exponential-like behaviour, and hence a finite correlation length may not exist on either side of the phase transition.
\end{theorem}
Overall, the conclusions outlined in this section indicate that spontaneous symmetry breaking in QFTs at finite temperature is significantly more complicated than in the vacuum case. The Goldstone mode continues to exist at all temperatures, but its properties are highly modified by its interactions with the thermal medium, which are non-universal and controlled by the microscopic dynamics of the theory.

\section{Goldstone evolution across the $\mathrm{U}(1)$ phase transition}
\label{goldstone_u1}

Given the various theoretical implications of Goldstone's theorem outlined in Sec.~\ref{Goldtone_QFT}, there is a strong motivation to test whether these characteristics are in fact realised in QFT models with a genuine finite-temperature phase transition. In particular, are thermal Goldstone modes present above the critical temperature $T_{c}$, and if so, do they have the characteristic structure of a massless thermoparticle? In Ref.~\cite{Lowdon:2025fyb} it was demonstrated in the $\mathrm{U}(1)$ complex scalar theory that such a thermal Goldstone mode is indeed present, but this study was restricted to a single temperature below and above $T_{c}$. Here we extend this analysis by following the presence of this mode for different temperatures across the transition region. 

\subsection{Thermal Goldstone correlators}
\label{thermal_correl}

To extract thermal Goldstone properties from the lattice one must first understand how these modes manifest themselves in the infinite-volume theory. In Ref.~\cite{Lowdon:2025fyb} it was shown that for a complex scalar theory with Euclidean two-point function $C(\tau,\vec{x})=\langle\phi(\tau,\vec{x})\phi^{\dagger}(0)\rangle_{\beta}$, a massless thermoparticle Goldstone mode, which shows up as $D_{\beta}^{G}(\vec{x})\delta(s)$ in the corresponding thermal spectral density, gives the following distinct contribution to the two-point function in the spatial direction:  
\begin{align}
C^{G}(0,\vec{x}) = \frac{\coth\left(\frac{\pi |\vec{x}|}{\beta} \right)}{4\pi \beta |\vec{x}| }D_{\beta}^{G}(\vec{x})  \xlongrightarrow{T\rightarrow 0}{} \frac{\alpha_{0}}{4\pi^{2}|\vec{x}|^{2}},
\label{Goldstone_T}
\end{align}
which approaches the standard form for the correlator of a massless particle state in the zero-temperature limit, with $\alpha_{0}$ a constant. An important point to note is that the spontaneous symmetry breaking condition in Eq.~\eqref{SSB_q} implies that the thermal Goldstone mode contributes to the $\langle j_{0}(x)\phi(0)\rangle_{\beta}$ correlation function. However, given that this mode exists, and both the current and field have the same quantum numbers, it must also give contributions to $\langle \phi(x)\phi^{\dagger}(0)\rangle_{\beta}$ as well. Although the damping factors of the Goldstone mode $D^{(+)}_{\beta}(\vec{x})$ and $D_{\beta}^{G}(\vec{x})$ which appear in these respective correlation functions are not the same, the dissipative effects must \textit{also} manifest themselves in the behaviour of $D_{\beta}^{G}(\vec{x})$, since the change in $D^{(+)}_{\beta}(\vec{x})$ below and above $T_{c}$ reflects a global property of the thermal medium itself. \\

\noindent
Another Euclidean correlation function which was important for the analysis in Ref.~\cite{Lowdon:2025fyb} was the spatial screening correlator $C(z) = \int  \!dx \, dy \, d\tau \, C(\tau,\vec{x})$. Given that the Goldstone mode has the form $D_{\beta}^{G}(\vec{x})\delta(s)$ in the thermal spectral density, it follows~\cite{Lowdon:2022xcl} that its contribution to the spatial screening correlator can be written
\begin{align}
C^{G}(z) = \frac{1}{2} \int^{\infty}_{|z|} \! dR \  D_{\beta}^{G}(R),
\label{CG_int}
\end{align}
and hence information about the damping factor can be extracted directly from spatial correlator data, in particular $D_{\beta}^{G}(|\vec{x}|=z)$ is proportional to the derivative of $C^{G}(z)$. Although it was not explicitly discussed in Ref.~\cite{Lowdon:2025fyb}, one can also investigate the temporal correlator $C(\tau) = \int  \!dx \, dy \, dz \, C(\tau,\vec{x})$, since this provides complementary information regarding the spectral properties of the theory. In particular, the validity of any spectral components extracted from $C(z)$ can be tested by comparing the corresponding predictions of $C(\tau)$ with the lattice data. This has been shown in both scalar theories~\cite{Lowdon:2024atn} and QCD~\cite{Lowdon:2022xcl,Bala:2023iqu} to provide a highly non-trivial test of the spectral components extracted from lattice data.

\subsection{Lattice analysis of the $\mathrm{U}(1)$ theory}

In this work we extend the analysis of the $\mathrm{U}(1)$ lattice scalar theory initiated in Ref.~\cite{Lowdon:2025fyb} by investigating a range of temperatures across the transition region. Here we briefly summarise the main characteristics of this model and the lattice setup\footnote{More details can be found in Ref.~\cite{Lowdon:2025fyb}.}. At finite temperature, the $\mathrm{U}(1)$ complex scalar field theory is known to possess two phases: a spontaneously-broken phase for $T<T_{c}$, and a symmetry-restored phase for $T>T_{c}$~\cite{Kapusta:2006pm}. For $T=0$ the vacuum expectation value of the scalar field is non-vanishing, in particular $|v|^{2}=\langle \phi \rangle\langle \phi^{\dagger} \rangle>0$. Here the model contains a massless Goldstone mode and a resonance-like excited mode. Above $T_{c}$ one finds that $|v|^{2}=0$, which implies that the global $\mathrm{U}(1)$ symmetry is restored. The lattice action of the model used in this analysis has the explicit form  
\begin{align}
S &= a^{4} \! \sum_{x \in \Lambda_{a}} \left[ \sum_{\mu} \left( \frac{1}{2} \Delta_{\mu}^{f} \phi^{*}(x)\Delta_{\mu}^{f}\phi(x)\right) + \frac{m_{0}^{2}}{2}\phi^{*}(x)\phi(x) + \frac{g_0}{4!}\left(\phi^{*}(x)\phi(x)\right)^{2} \right],
\label{Latt_action}
\end{align}
where $\Delta_{\mu}^{f}$ is the lattice forward derivative. We keep the lattice spacing $a$ fixed throughout in order to avoid complications due to the possible triviality of the theory. The action is defined on a spatially symmetric space-time volume $L^{3}\times L_\tau$, where $L_{\tau}= aN_{\tau}$ and $L = aN_{s}$, and the temperature of the system is defined as $T= (aN_{\tau})^{-1}$. The vacuum phase of the theory depends on the specific value of the bare parameters $m_{0}$ and $g_{0}$. The specific choice $(am_{0},g_{0})=(0.297 i,0.85)$ leads to a theory with a symmetry-broken vacuum, while allowing the exploration of temperature regimes both below and above $T_{c}$. The vacuum expectation value $|v|$ at zero temperature sets the physical scale of the theory, in which all dimensionful quantities can be expressed. For our choice of bare parameters the UV cutoff satisfies $\Lambda/|v|=\pi(a|v|)\approx 36$, where $|v|$ is extrapolated from the coldest lattice with $N_{\tau}=32$. The largeness of $\Lambda/|v|$ implies that cutoff effects for correlators beyond a few lattice spacings are very small, and hence the continuum formulae discussed in the previous section can be reliably applied to study the mid and long-range behaviour of our lattice correlators. \\

\noindent
Determining the phase of the lattice $\mathrm{U}(1)$ theory is non-trivial since spontaneous symmetry breaking can only occur in an infinite volume. A careful analysis therefore requires an $L\rightarrow \infty$ extrapolation of the lattice results. There are different approaches for estimating $|v|$ from lattice data~\cite{Neuberger:1987fd,Neuberger:1987zz}, but most of these make use of the fact that the finite spatial-volume Euclidean two-point function $C_{L}(\tau,\vec{x})$ satisfies the condition 
\begin{align}
\lim_{|\vec{x}| \rightarrow \infty}\lim_{L\rightarrow \infty}C_{L}(\tau,\vec{x}) \longrightarrow |v|^{2}.
\end{align}
In this analysis we use the definition   
\begin{align}
|v|^{2} = \lim_{L\rightarrow \infty}  C_{L}(0,|\vec{x}|=L/2),
\label{vev_def}
\end{align}  
and hence an estimate of $|v|^{2}$ can be obtained by extrapolating the $L\rightarrow \infty$ behaviour using a range of correlators at sufficiently large values of $N_{s}$. \\

\begin{figure}[t!]
\centering
\includegraphics[width=0.6\textwidth]{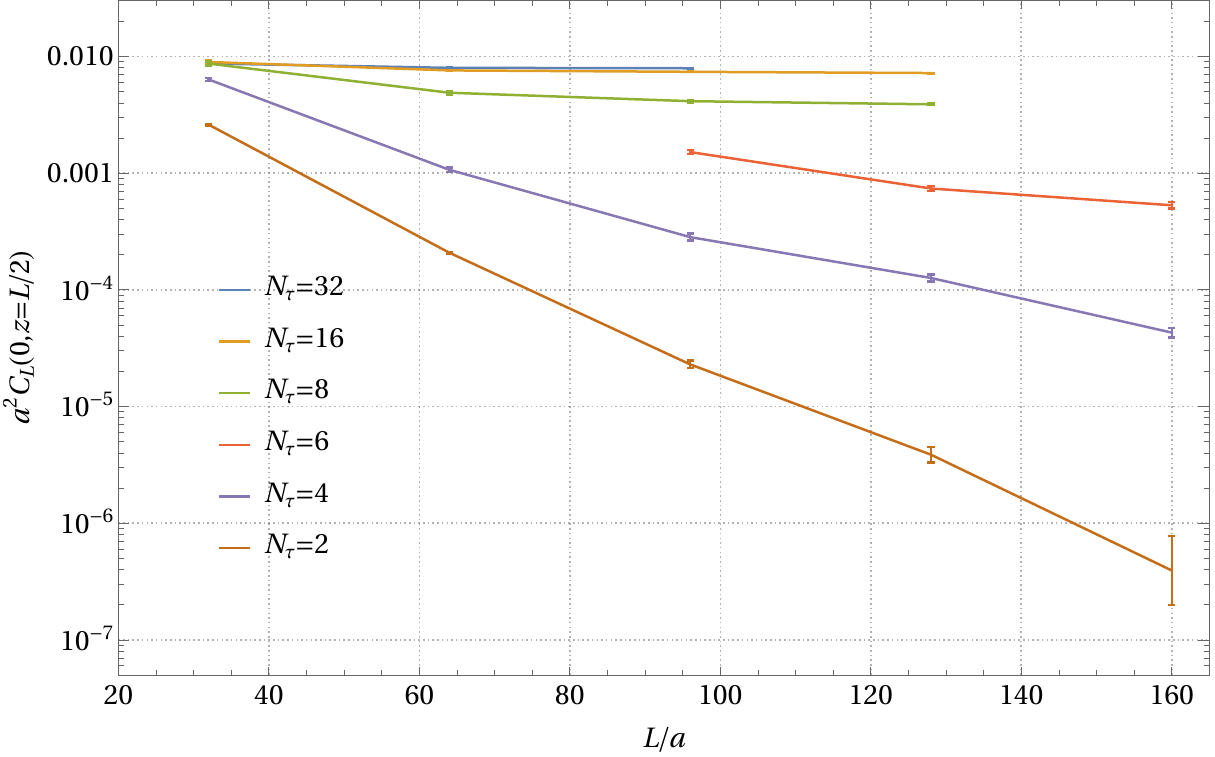}
\caption{$C_{L}(0,z=L/2)$ as a function of $L/a$ for different values of $N_{\tau}$.}
\label{C_L_Nt}
\end{figure}

\noindent
In Ref.~\cite{Lowdon:2025fyb} we analysed volumes with $N_{s} \geq 32$ at $N_{\tau}=2$ and $N_\tau=32$, in order to deeply probe the system in the symmetric and broken phases, respectively. For this analysis, we include additional intermediate temperatures at $N_{\tau}=4, 6, 8, 16$. Fig.~\ref{C_L_Nt} shows the finite-volume correlator data for $C_{L}(0,z=|\vec{x}|=L/2)$ as a function of $L$ for each value of $N_{\tau}$.
For $N_{\tau}=8,16,32$ the correlator converges towards a non-zero result for increasing volume, whereas for $N_{\tau}=2,4,6$ the correlator approaches a vanishing value in the infinite-volume limit. From Eq.~\eqref{vev_def} this suggests that the infinite-volume system is consistent with being in the symmetry-broken phase for $N_{\tau}=8,16,32$, and the symmetry-restored phase for $N_{\tau}=2,4,6$. The inclusion of the additional $N_{\tau}$ points has therefore enabled the system to be investigated in more detail both below and above $T_{c}$. In the following two subsections we will summarise the results of the analyses in both of these phases.

\subsubsection{Broken phase}
\label{broken_phase}

\begin{figure}[t!]
\centering
\includegraphics[width=0.45\textwidth]{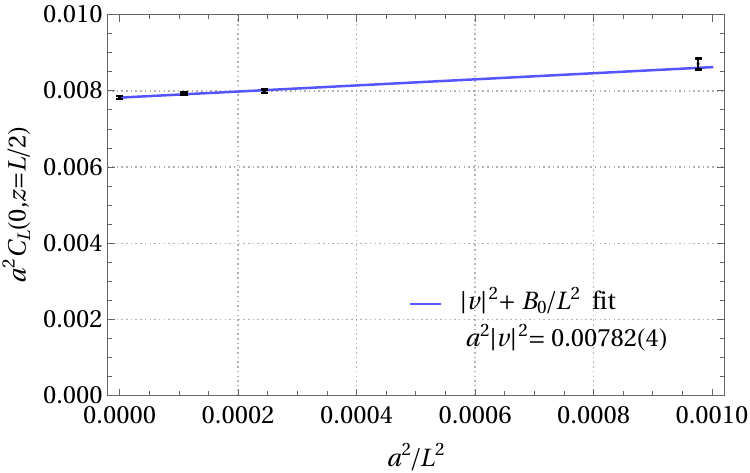}\
\includegraphics[width=0.45\textwidth]{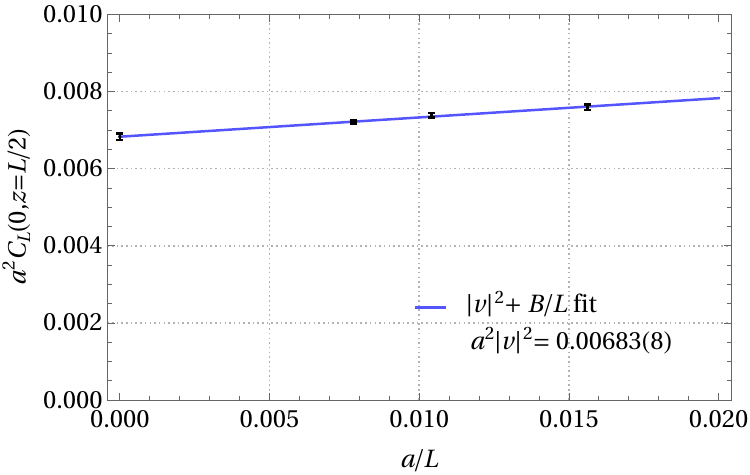}    
\includegraphics[width=0.45\textwidth]{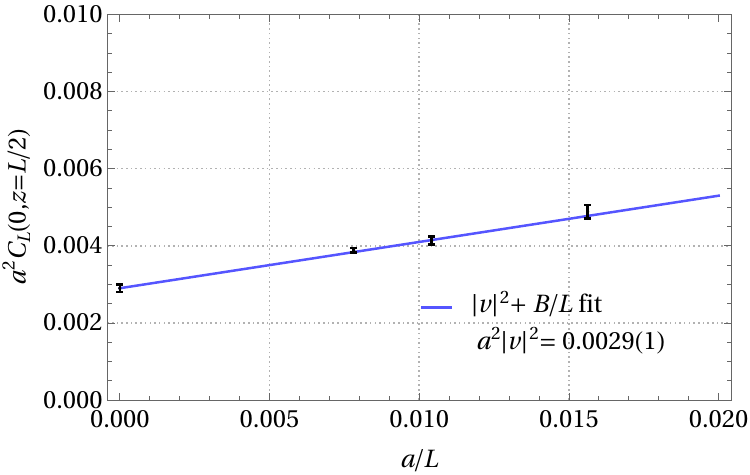}
\caption{Infinite-volume extrapolation of $a^{2}C_{L}(0,L/2)$ for $N_{\tau}=32$ (top left), $N_{\tau}=16$ (top right), and $N_{\tau}=8$ (bottom).}
\label{CL_Nt}
\end{figure}

From Fig.~\ref{C_L_Nt} and the extrapolation in Eq.~\eqref{vev_def}, one can immediately see that the data points at $N_{\tau}=8,16,32$ are qualitatively consistent with being in the broken phase. In Ref.~\cite{Lowdon:2025fyb} a quantitative estimate for the non-vanishing value of $|v|^{2}$ was made by fitting $C_{L}(0,z=L/2)$ to the functional form $|v|^{2} + B_{0}/L^{2}$, resulting in the infinite-volume extrapolation: $a^{2}|v|^{2} =  0.00782(4)$. This was motivated by the fact that the $C_{L}(0,z)$ lattice correlator data at different volumes ($N_{s}= 32, 64, 96$) was consistent with the finite-volume massless particle-like form 
\begin{align}
C_{L}(0,z) = c_{0} + b_{0} \left[ \frac{1}{z^{2}} + \left\{ z \rightarrow (L-z)\right\} \right], 
\label{vac_fit}
\end{align}
where fits were performed over a wide spatial range $[z_{\text{min}},L/2]$, and shown to be stable under large variations of $z_{\text{min}}$. In order to analyse the higher temperature data at $N_{\tau}=8$ and $16$ we first repeated the same procedure as for $N_{\tau}=32$ by using Eq.~\eqref{vac_fit}. We found that in both cases the correlator data could not be consistently described with this behaviour, since the extracted fit parameters were highly sensitive to $z_{\text{min}}$. This indicates that temperature effects start to play a significant role for $N_{\tau}<32$. Since the thermal Goldstone mode is expected to have the spatial correlator structure in Eq.~\eqref{Goldstone_T} at all temperatures, we performed fits with the following finite-volume ansatz
\begin{align}
C_{L}(0,z)  = c_{L} +  b_{L} \left[\frac{\coth\left(\frac{\pi z}{\beta} \right)}{z} + \left\{z \rightarrow (L-z)\right\}  \right], 
\label{C_L_broken}   
\end{align}
which describes the situation where temperature effects are significant enough to result in modifications from the $\coth$ pre-factor, but do not lead to a significant damping in the correlator, and hence $D_{\beta}^{G}(\vec{x}) \approx \text{const}$. Upon performing fits of the $N_{\tau}=8$ and $16$ correlator data with Eq.~\eqref{C_L_broken} we found that the data was indeed highly consistent with this functional form, with very little sensitivity of the fit parameters to $z_{\text{min}}$ over a wide range of values. In order to further assess the robustness of these fits we also tested range of parametrisations of the form: $c_{L} + b_{L} \left[ z^{-n} + \{ z \rightarrow (L-z) \} \right]$ with $n \neq 1$, but found that none of these were able to consistently describe the data. Since $\coth(\pi z/\beta)\sim \text{const}$ at large values of $z$ for $N_{\tau}=8$ and $16$, the consistency of Eq.~\eqref{C_L_broken} suggests that the $C_{L}(0,z=L/2)$ data should be well-described by the functional form $|v|^{2} + B/L$, and hence one can use this parametrisation to perform an infinite-volume extrapolation of $|v|^{2}$. Using the volumes $N_{s}=64, 96, 128$ we obtained significant fits at both of these temperatures, resulting in the infinite-volume extrapolations: $a^{2}|v|^{2}(N_{\tau}=16)=0.00683(8)$, and $a^{2}|v|^{2}(N_{\tau}=8)=0.0029(1)$. The extrapolation fits for $N_{\tau}=8, 16, 32$ are displayed in Fig.~\ref{CL_Nt}. This analysis confirms that at $N_{\tau}=8, 16, 32$ the system is in the symmetry-broken phase, with $|v|^{2}$ decreasing successively with increasing temperature, and that the scalar correlation function is dominated by a Goldstone mode which experiences weak dissipative effects, as indicated by the conclusions of Sec.~\ref{T_goldstone}.   

\subsubsection{Symmetry-restored phase}
\label{restored_phase}

In contrast to $N_{\tau}=8,16,32$, from Fig.~\ref{C_L_Nt} one can see that for $N_{\tau}=2,4,6$ the $C_{L}(0,z=L/2)$ data points are rapidly decreasing as a function of $L$. In fact, this decrease approaches a linear-like behaviour, which given the logarithmic $|v|^{2}$ axis scale indicates that the correlator decays exponentially, and hence vanishes in the $L\rightarrow \infty$ limit. Due to Eq.~\eqref{vev_def} the $N_{\tau}=2,4,6$ points are therefore consistent with being in the symmetry-restored phase. In Ref.~\cite{Lowdon:2025fyb}, lattice correlator data for $C_{L}(0,z)$ was fitted at different volumes ($N_{s}= 32, 64, 96$) for $N_{\tau}=2$ using the ansatz
\begin{align}
C_{L}(0,z)  =  b_{L}  \left[\tfrac{\coth\left(\frac{\pi z}{\beta} \right)}{z} e^{-\gamma_{L} z} + \left\{z \rightarrow (L-z)\right\} \right], 
\label{C_L}    
\end{align}
and found to be consistent with Eq.~\eqref{C_L} over a wide fit range $[z_{\text{min}},L/2]$. This parametrisation was motivated by the general structure of the spatial thermal Goldstone correlator in Eq.~\eqref{Goldstone_T}, and the fact that the spatial screening correlator $C(z)$ could be consistently fitted over the full data range $[0,L/2]$ at each volume ($N_{s}=64, 96, 128$) using the finite-volume exponential parametrisation 
\begin{align}
C_{L}(z) = d_{L} \left[ e^{-m_{L} z} +  \left\{ z \rightarrow (L-z)\right\} \right], 
\label{Cz_L}  
\end{align}      
which in light of Eq.~\eqref{CG_int} implies that the Goldstone damping must have a pure exponential form
\begin{align}
    D_{\beta}^{G}(\vec{x}) = \alpha \, e^{-\gamma |\vec{x}|}.
\end{align}
A highly non-trivial test that the thermal Goldstone mode does indeed have an exponentially-damped massless thermoparticle structure is that the screening mass $m_{L}$ and damping factor exponent $\gamma_{L}$ must converge for $L \rightarrow \infty$, which was found to be the case in Ref.~\cite{Lowdon:2025fyb}. \\

\begin{figure}[t!]
  \begin{minipage}[t]{.5\textwidth}
\centering
\renewcommand{\arraystretch}{1.15}
\small
  \vfill
\begin{tabular}{|c|c|c|c|} 
\hline
\rule{0pt}{3ex}
$N_{s}^{3} \times N_{\tau}$ &   $d_{L}/a$  &  $am_{L}$       \\[0.5ex]
\hhline{|=|=|=|}
$96^{3} \times 2$           &   9.838(5)   &  0.10156(7)  	 \\
\hline
$128^{3} \times 2$          &   9.874(3)   &  0.10121(4)   	 \\
\hline
$160^{3} \times 2$          & 	9.857(3)   &  0.10094(3)     \\
\hline 
$96^{3} \times 4$           &   26.16(8)   &  0.0426(1)      \\
\hline 
$128^{3} \times 4$          &   27.20(2)   &  0.03797(3)     \\
\hline 
$160^{3} \times 4$          & 	26.00(4)   &  0.04054(7)     \\
\hline
$96^{3} \times 6$           &   82.12(1)   &  0.015481(4)    \\
\hline
$128^{3} \times 6$          &   82.10(2)   &  0.014457(5)    \\
\hline
$160^{3} \times 6$          & 	96.37(3)   &  0.012258(5)    \\
\hline
\end{tabular} 
\end{minipage}
  \begin{minipage}[t]{.5\textwidth}
  \vfill
    \centering
    \includegraphics[width=1\linewidth]{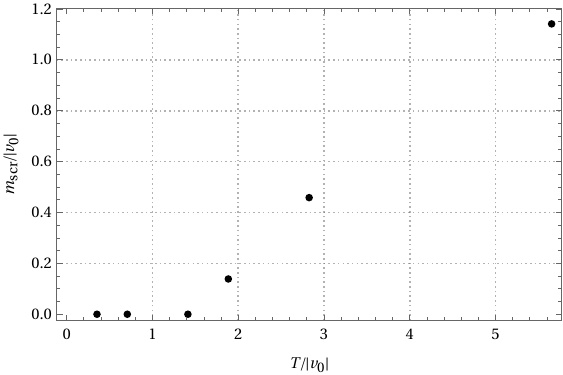} 
\label{Mscr_Nt}
  \end{minipage}
\caption{Fit parameter values of $d_{L}/a$ and $am_{L}$ obtained using the fit ansatz in Eq.~\eqref{Cz_L} for $N_{\tau}=2,4, 6$ at different volumes (left), and the screening mass $m_{\text{scr}}$ at $N_{s}=160$ versus temperature normalised to the field expectation value $|v_{0}|$ on the coldest lattice $N_{\tau}=32$ (right). The error bars in the plot are smaller than the symbol size.}
\label{m_L} 
\end{figure}

\noindent
An important question is whether the lattice data at smaller temperatures within the symmetry-restored phase are also consistent with this specific thermoparticle structure. To address this question we performed an identical analysis approach for $N_{\tau}=4,6$ over a wide range of different volumes: $N_{s}=32,64,96,128,160$ for $N_{\tau}=4$, and $N_{s}=96,128,160$ for $N_{\tau}=6$. We chose $N_{s} \geq 96$ for $N_{\tau}=6$ because we observed significant finite-volume corrections at smaller volumes due to its closer proximity to the phase transition. We found that for both $N_{\tau}=4$ and $N_{\tau}=6$ the data was highly consistent with the parametrisations in Eqs.~\eqref{C_L} and~\eqref{Cz_L}, and that $m_{L}$ and $\gamma_{L}$ also tended towards a common value as the volume increased. The values of the screening masses $m_{L}$ and coefficients $d_{L}$ for $N_{s} \geq 96$ obtained for $N_{\tau}=2, 4, 6$ are listed in the table of Fig.~\ref{m_L}. In the plot of Fig.~\ref{m_L} we also display the temperature dependence of the screening masses $m_{\text{scr}}$ on the largest lattice volume $N_{s}=160$. Since the field expectation value in the vacuum broken-symmetry phase $|v_{0}|$ sets the physical scale of the system, we normalise both $m_{\text{scr}}$ and $T$ by $|v_{0}|$, using the infinite-volume extrapolated value of $|v|$ on the coldest lattice $N_{\tau}=32$ to approximate $|v_{0}|$. For completeness, in Fig.~\ref{m_L} we also plot the screening masses obtained in the broken phase at $N_{\tau}=8, 16, 32$, which are consistent with zero at the current accuracy. \\

\begin{figure}[t!]
\includegraphics[width=0.32\textwidth]{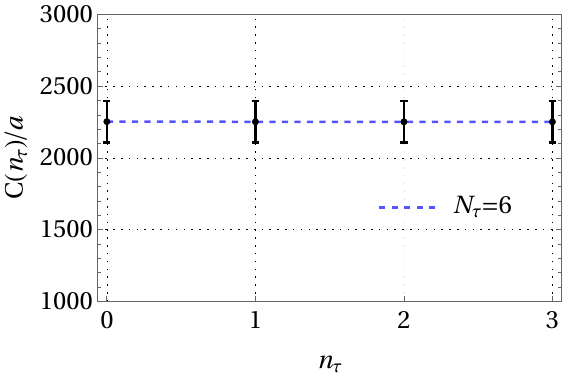}
\includegraphics[width=0.32\textwidth]{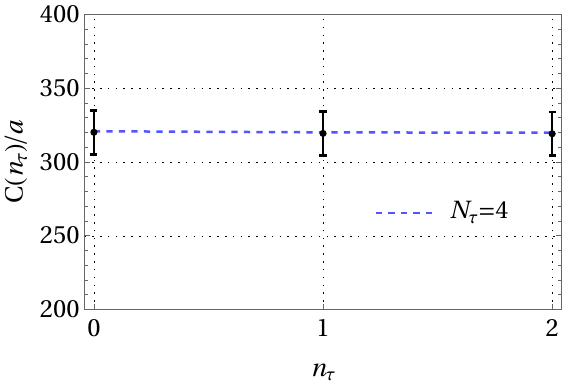}
\includegraphics[width=0.32\textwidth]{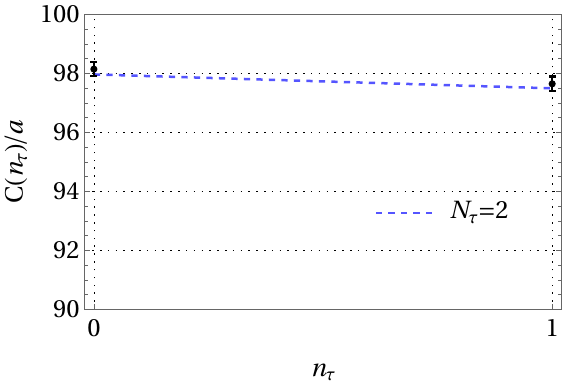}
\caption{Thermal Goldstone temporal correlator prediction (blue dashed line) and data (black points) for $L/a=160$ at $N_{\tau}=6$ (left), $N_{\tau}=4$ (middle), and $N_{\tau}=2$ (right).}
\label{Wt_Nt}
\end{figure}

\noindent
These results strongly indicate that the spatial two-point function is dominated by a single thermal Goldstone component with the exponentially-damped massless thermoparticle structure
\begin{align}
C^{G}(0,\vec{x})  =  \frac{\coth\left(\frac{\pi |\vec{x}|}{\beta} \right)}{4\pi \beta |\vec{x}| }\alpha \, e^{-\gamma |\vec{x}|}.
\label{CG}
\end{align} 
This implies that the spectral function arising from this component $\rho_{G}(\omega,\vec{p})$ has the form~\cite{Lowdon:2025fyb} 
\begin{align}
\rho_{G}(\omega,\vec{p}) = \frac{4\alpha \, \omega \gamma }{(\omega^{2}-|\vec{p}|^{2}-\gamma^{2})^{2} + 4\omega^{2}\gamma^{2}}.
\label{rhoG}
\end{align} 
As outlined in Sec.~\ref{thermal_correl}, another way to test the consistency of these findings is to compare these results with the data from the temporal correlator $C(\tau)$, since $C(\tau)$ has a significantly different dependence on the spectral function than $C(z)$~\cite{Lowdon:2024atn,Lowdon:2022xcl,Bala:2023iqu}. If the thermal Goldstone mode dominates the spectral function it follows that the temporal correlator can be written 
\begin{align}
C(\tau) = \int_{0}^{\infty} \frac{d\omega}{2\pi} \frac{\cosh\left[\left(\frac{\beta}{2}-|\tau| \right)\omega\right] }{\sinh\left(\frac{\beta}{2}\omega\right)} \,\rho_{G}(\omega,\vec{p}=0). 
\label{GcorrT}
\end{align}
Combining Eqs.~\eqref{CG_int} and~\eqref{CG}, the damping factor coefficient $\alpha$ is related to the infinite-volume limit of the spatial correlator coefficient $d_{L}$ and screening mass $m_{L}$ via
\begin{align}
\alpha = \lim_{L\rightarrow \infty} 2 d_{L} m_{L}.
\end{align}
Together with the relation $\gamma = \lim_{L\rightarrow \infty} m_{L}$ one can then use the parameters in the table of Fig.~\ref{m_L} at the largest volume $L/a=160$, and Eqs.~\eqref{rhoG} and~\eqref{GcorrT}, to predict the form of $C(\tau)$, and compare this with the corresponding data. This comparison is displayed in Fig.~\ref{Wt_Nt} for $N_{\tau}=2,4,6$. 
In each case\footnote{For $N_{\tau}=6$ finite-volume corrections still have an effect because of the smallness of $am_{L}$. In order to take this into account we used the perturbatively-inspired finite-volume-corrected relation: $\alpha_{L} = 2 d_{L} m_{L}(1-e^{-Lm_{L}})$.}, we found that the $C(\tau)$ prediction from the thermal Goldstone mode was consistent with the data within errors, which is further evidence that these modes are indeed present, and dominate the correlators up to the highest temperatures studied here. \\

\begin{figure}[t!]
\centering
\includegraphics[width=0.49\textwidth]{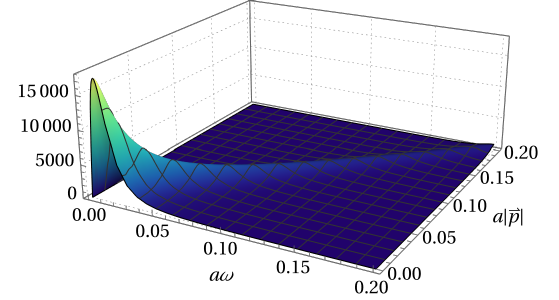}
\includegraphics[width=0.49\textwidth]{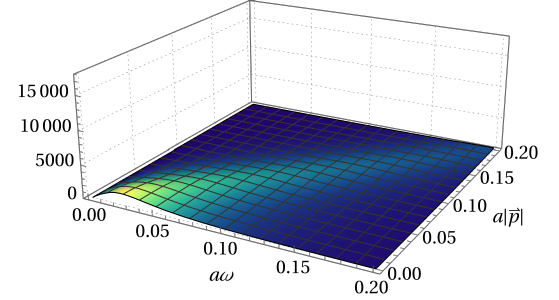}
\includegraphics[width=0.49\textwidth]{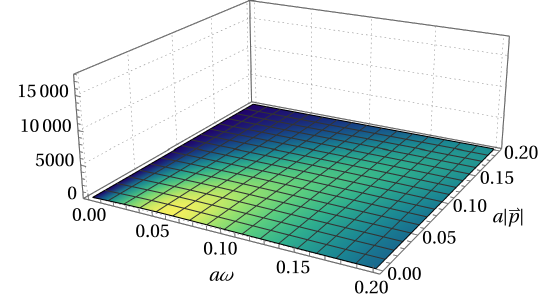}
\caption{Evolution of the Goldstone spectral function $\rho_{G}(a\omega,a|\vec{p}|)/a^{2}$ with increasing temperature for $L/a=160$ at $N_{\tau}=6$ (top left), $N_{\tau}=4$ (top right), and $N_{\tau}=2$ (bottom).}
\label{rho_G_Nt}
\end{figure}

\noindent
In order to visualise the spectral structure of the thermal Goldstone mode in the symmetry-restored phase, in Fig.~\ref{rho_G_Nt} we plot $\rho_{G}(\omega,\vec{p})$ for $N_{\tau}=2,4,6$ on the largest volume $L/a=160$. With the vertical axis scale set to the same range in each plot, one can immediately see how the Goldstone mode evolves as a function of temperature. It starts with a very narrow width close to the transition at $N_{\tau}=6$, and is densely concentrated near the mass shell $p^{2}=0$. As the temperature is raised at $N_{\tau}=4$, and then further at $N_{\tau}=2$, the amplitude of the spectral peak is significantly reduced, and the spectral function width broadens. The state therefore loses its localised particle-like behaviour above $T_{c}$ due to the increasingly strong dissipative effects it experiences in the hot thermal medium.   

\subsection{Spectral characteristics across the $\mathrm{U}(1)$ phase transition}
 
The main conclusion from the analysis of Secs.~\ref{broken_phase} and~\ref{restored_phase} is that the connected scalar correlator in the spatial direction $C_{c}(0,\vec{x})=C(0,\vec{x})-|v|^{2}$ has the following functional form
\begin{align}
C_{c}(0,\vec{x}) = \frac{\coth\left(\frac{\pi |\vec{x}|}{\beta} \right)}{4\pi \beta |\vec{x}| }D_{\beta}^{G}(\vec{x}), \quad 
D_{\beta}^{G}(\vec{x}) \approx \left\{ 
\begin{array}{ll}
      1, & N_{\tau} = 8,16,32  \\
      e^{-\gamma |\vec{x}|}, & N_{\tau} = 2,4,6 
    \end{array}   \right.  
\label{Cz_concl}   
\end{align}
with $N_{\tau} =8,16,32$ in the broken phase, and $N_{\tau}=2,4,6$ in the symmetry-restored phase. This is consistent with the correlation function being dominated by a single massless thermoparticle state both below and above the phase transition, with the damping experienced by this state changing discontinuously as the transition is crossed. Since the screening mass $m_{\text{scr}}$ entirely controls the damping of the thermal Goldstone mode, this transition can be visualised in the plot of Fig.~\ref{m_L}. An interesting observation from Fig.~\ref{m_L} is that $m_{\text{scr}}$ approaches the physical scale of the system $|v_{0}|$ on the hottest lattice, and the temperature-dependence of $m_{\text{scr}}$ appears to approach a linear-like behaviour in the symmetry-restored phase. In contrast to the conventional understanding~\cite{Kapusta:2006pm}, the Goldstone mode does not behave like a vacuum massless particle in the broken phase, except at $T=0$. Although we observe negligible damping in the symmetry-broken regime, we expect that with higher precision data non-trivial damping effects could be extracted, particularly close to the transition temperature. The findings summarised in Eq.~\eqref{Cz_concl} are directly in line with the results of Sec.~\ref{Goldtone_QFT}, namely that thermal phase transitions are characterised by a transition from a regime in which the Goldstone mode experiences weak dissipation, to one in which the dissipation is strong. Although this condition directly relates to the damping factor $D^{(+)}_{\beta}(\vec{x})$ of the current-field correlator, it makes sense that the damping factor $D_{\beta}^{G}(\vec{x})$ of the scalar correlator is also sensitive to the dissipation change across the transition, since this reflects a global property of the medium. 

\subsection{Contrast with existing studies} 

Much of the emphasis in the literature has centred around understanding the characteristics of thermal Goldstone modes in the spontaneously broken regime $T<T_{c}$. An example of particular physical importance is the pions in QCD, which correspond to the Goldstone modes associated with spontaneous chiral symmetry breaking in the limit of vanishing quark masses. For $T<T_{c}$ it has been suggested that the Goldstone pions have a quasi-particle like structure~\cite{Shuryak:1990ie,Pisarski:1996mt,Son:2001ff,Son:2002ci}, and that for sufficiently small momenta the real part of their dispersion relations can be extracted from Euclidean correlation functions~\cite{Son:2001ff,Son:2002ci,Brandt:2014qqa,Krasniqi:2024kwm}. Although significantly less focus has been given to the $T>T_{c}$ case, other studies have also attempted to understand what happens in this regime, including more recently in Refs.~\cite{Krasniqi:2024kwm,Florio:2021jlx,Florio:2025zqv}, where it is argued that Goldstone excitations behave like non-propagating diffusive modes.  \\

\noindent
The theoretical results outlined in Sec.~\ref{TGen} and the corresponding lattice evidence detailed in this section contrast significantly with these existing studies, particularly for $T>T_{c}$, where we have demonstrated that thermal Goldstone modes continue to exist as propagating massless thermoparticles, even though the symmetry is restored. Most likely, this difference to the previous literature arises from the basic assumptions made in the earlier analyses. A common feature of many of the existing studies is the use of a hydrodynamical approach. As outlined in Ref.~\cite{Son:2001ff}, a fundamental assumption of hydrodynamics is that the correlation functions contain only pole-like singularities. Whilst this is certainly true for $T=0$, an important question is whether this remains the case when $T>0$. In Ref.~\cite{Bros:2003zs} this question was investigated using the same non-perturbative QFT approach used to establish the conclusions in Sec.~\ref{TGen}. It was shown that thermal correlation functions can in fact have a much broader class of singularities than their vacuum counterparts, and that these singularities will in general be more complicated than simple poles. This indicates that a hydrodynamical description may well fail to capture the effects of certain types of thermal QFT excitations, including thermoparticles. If so, this explains the discrepancies with the existing literature.

\section{Conclusions}
\label{concl} 

How spontaneous symmetry breaking manifests itself for thermal states is central to understanding the characteristics of phase transitions at finite temperature. A particularly important question is what happens to the Goldstone bosons that exist in the vacuum theory as the temperature is increased. By investigating the $\mathrm{U}(1)$ complex scalar field theory on the lattice, in Ref.~\cite{Lowdon:2025fyb} we confirmed the findings of Ref.~\cite{Bros:1996kd} that thermal Goldstone modes can continue to exist at high temperatures, even above the critical temperature $T_{c}$ where the symmetry is restored, and have the properties of screened massless particle-like excitations, so-called thermoparticles. In this work, we further explored the analytic results of Ref.~\cite{Bros:1996kd}, showing that the thermal phase properties of QFTs can be entirely characterised by the dissipative properties experienced by the thermal Goldstone mode. Weak dissipation of the mode implies that the system is in a broken phase, and strong dissipation ensures that the symmetry is restored, which is consistent with the physical picture that high-temperature effects can destroy the long-range order of a system. By continuing the lattice analysis of the $\mathrm{U}(1)$ model at several different temperatures below and above $T_{c}$, we found that the thermal Goldstone mode does indeed experience a discontinuous change in its dissipative properties, which strongly supports this picture. The damping of thermal Goldstone modes provides a new way in which to characterise phase transitions in QFTs at finite temperature, and has consequences for a range of physical phenomena, including the high-temperature properties of quasi-particle excitations in condensed matter systems, and chiral symmetry restoration in QCD.

\section*{Acknowledgements}
The authors acknowledge support by the Deutsche Forschungsgemeinschaft (DFG, German Research Foundation) through the Collaborative Research Center CRC-TR 211 ``Strong-interaction matter under extreme conditions'' -- Project No. 315477589-TRR 211. O.~P.~also acknowledges support by the State of Hesse within the Research Cluster ELEMENTS (Project ID 500/10.006).

\bibliographystyle{JHEP}
 
\bibliography{refs}

\end{document}